\documentclass[12pt]{amsart}
\usepackage{amsmath}
\usepackage{color}
\usepackage{amsfonts}
\usepackage{amssymb}
\usepackage{epsfig}
\usepackage{times}
\usepackage{braket}
\usepackage{shadow}
\usepackage{fancybox}
\usepackage{colordvi}
\usepackage{graphics}
\usepackage{epsf,graphics,mathrsfs,yfonts,eufrak,simplewick}
\usepackage{psfrag}
 \usepackage[enableskew]{youngtab}       

\newcommand{\be}{\begin{eqnarray*}}
\newcommand{\en}{\end{eqnarray*}}
\newcommand{\ba}{\begin{array}}
\newcommand{\ea}{\end{array}}
\newcommand{\bea}{\begin{eqnarray}}
\newcommand{\ena}{\end{eqnarray}}
\def\bel{\begin{eqnarray}}
\def\enl{\end{eqnarray}}
\def\({\left(}
\def\){\right)}
\def\[{\left[}
\def\]{\right]}
\newcommand{\half}{\textstyle{\frac 1 2 }}

\newcommand{\nn}{\nonumber}
\newcommand{\slt}{\mathfrak{sl}_2}

\newcommand{\Tr}{{\rm Tr}}

\newcommand{\la}{\lambda}

\newcommand{\al}{\alpha}

\newcommand{\s}{\sigma}

\begin{document}

\begin{title}[New exact results on density matrix for XXX spin chain]{
New exact results on density matrix for XXX spin chain}

\end{title}

\author{T.~Miwa and  F.~Smirnov}
\address{Institute for Liberal Arts and Sciences, Kyoto University}\email{tmiwa@kje.biglobe.ne.jp}
\address{FS\footnote{Membre du CNRS}: 
${}^{1}$ Sorbonne Universit\'e, UPMC Univ Paris 06\\ CNRS, UMR 7589, LPTHE\\F-75005, Paris, France}\email{smirnov@lpthe.jussieu.fr}

\begin{abstract}
Using the fermionic basis we obtain the expectation values of all 
$\slt$-invariant and $C$-invariant
local operators
on 10 sites for the anisotropic six-vertex model on a cylinder with generic Matsubara 
data. This is equivalent to the generalised Gibbs ensemble for the XXX spin chain.
In the case when the $\slt$ and $C$ symmetries are not broken this computation
is equivalent to finding the entire density matrix up to 10 sites.
As application, we compute the entanglement entropy without and with 
temperature, and compare the results with CFT predictions. 
 \end{abstract}

\maketitle

\section{Introduction}

Since remarkable works by Boos and Korepin \cite{bk} it became clear that
all the expectation values of local operators for XXX antiferromagnet
must be expressible in terms of  values $\zeta$-function at odd positive
integer arguments. This statement was proved in the paper \cite{XXX}.
Methods of this paper were used by Takahshi {\it et al.} \cite{takahashi} to compute the
correlation functions of spins up to $8$ sites, and the density
matrix up to $6$ sites. The latter computation allowed one to find the
entanglement entropy.

In the paper \cite{HGSIII} the computation of expectation values
is put in rather general framework. The main ingredient used in this paper is the fermionic basis. It is shown that this basis allows one to compute the
expectation values on a cylinder with arbitrary Matsubara data. This
circumstance was used in the 
paper \cite{corr} in order to find an analog of OPE on the lattice: the coefficients
expressing a local operator in terms of the fermionic basis. The expectation
values for the latter are simple. 

In the present paper we apply the methods of \cite{corr} to the expectation values of all the $\slt$-invariant and
$C$-invariant operators
for subchains of up tp $10$ sites. Namely, we decompose all of
them in the fermionic basis. Then the expectation values for any Matsubara data are
easy to compute. 

When the $\slt$-symmetry and {$C$-invariance} are not broken by the Matsubara
eigenvector (anti-ferromagnet with temperature, but without magnetic field) our
results are sufficient to derive entire density matrix. 
As application we compute the entanglement entropy for
zero temperature and for small temperatures different from zero, and compare
the results with the CFT predictions. The agreement is good, so, $n=10$ seems to be already a large number. 

The paper consists of six sections and one Appendix. In Section 2 we give some
information about the fermionic basis. In Section 3 we explain how to compute
efficiently the expectation values of operators with small Matsubara lattices.
The computation is based on Slavnov formula \cite{Slavnov} for {scalar} product and some
basic formulae of QISM \cite{FST,BIK}. We solve the combinatorial problem
of expressing the results in terms of Schur polynomials. Section 4  summarises the
computational procedure. In section 5 we compute the density matrix and
entanglement entropy for zero temperature. In Section 6 we explain
how to compute efficiently at non-zero temperature the basic object of our method which is the function
{$\omega$.}
In Section 7 we present results for the entanglement entropy at {small temperatures different from zero} and compare them with the CFT prediction. In Appendix
we present the eigenvalues of the density matrix at zero temperature with the precision $10^{-11}$.

\section{Fermionic basis}\label{fermionic}

Fermionic basis for the case of $\slt$-invariant and $C$-invariant operators
is explained in details in \cite{corr}. So, we shall be brief here. 
We have two sets of fermionic  operators $b_j,\ b^*_j$, $c_j,\ c_j^*$, ($j=1,2,3,\cdots$)
with canonical commutation relations,
and use notations  $b^*_J$, $c_J^*$ for products, $J$ being a strictly ordered
{multi-}index $\{j_1,\cdots,j_k\}$.
For two {multi-}indices of the same length
we write $I\preccurlyeq J$ if $i_p\le j_p$ for all $p$. We denote by {$|I|$} the sum of elements in $I$. 
Our fermionic  operators act on the space of local fields, role of vacuum is played by the unit
operator $\mathrm{I}$.
Consider the
space $\mathfrak{H}^{(n)}$ with the basis
\begin{align}b^*_Ic_J^*\cdot \mathrm{I},\label{base}\end{align}
with $\#(I)=\#({J}){\le [n/2]}$, $\max(I\cup J)\le n$, $|I|+|J|\equiv 0\ (\mathrm{mod}\ 2)$, $I\preccurlyeq J$.
Define the operators
\begin{align}
&Q_m=\sum_{j=1}^{m-1}c_jb_{m-j}\,,\quad  2,3,\cdots\,;\qquad
M=\sum\limits_{i=1}^\infty c^*_ib_i\,.\nn
\end{align}
Introduce the space  $\widetilde{\mathfrak{H}}^{(n)}$ defined as above with conditions $|I|+|J|\equiv 0\ (\mathrm{mod}\ 2)$, $I\preccurlyeq J$ lifted.
The operators $Q_m$ act from  $\mathfrak{H}^{(n)}$ to $\widetilde{\mathfrak{H}}^{(n)}$. The operator
$M$ acts from the space  $\widetilde{\mathfrak{H}}^{(n)}_2$ (space of charge $2$),  span by
the vectors \eqref{base} with $\#(I)+1=\#({J})-1{\le [n/2]}$, $\max(I\cup J)\le n$ , to
$\widetilde{\mathfrak{H}}^{(n)}$. We define the subspace 
${\mathfrak{V}}^{(n)}$ of ${\mathfrak{H}}^{(n)}$ 
by {
$$\mathfrak{V}^{(n)}=\{v\in\mathfrak{H}^{(n)}\mid Q_mv\in M\widetilde{\mathfrak{H}}^{(n)}_2
\ \text{for}\ \ m=n+1,n+2,\cdots\,\}.$$
}
It is easy to see that  $Q_m{\mathfrak{V}}^{(n)}=0$ for $m>2n-1$  , so the actual
number of requirements is finite.

Denoting basis of ${\mathfrak{V}}^{(n)}$ by $v_\al$ we have {$F=||F_{\al,\{I,J\}}||$, the
first one} of several matrices used below:
$$v_\al=\sum\limits_{ {\#(I)=\#({J}), \max(I\cup J)\le n}\atop{ |I|+|J|\equiv 0\ (\mathrm{mod}\ 2),
I\preccurlyeq J}}F_{\al,\{I,J\}}\ b^*_Ic_J^*\cdot \mathrm{I}\,.$$

From now on we shall demonstrate complexity of computation by {the most difficult case to be
considered in this paper, which is $n=10$.}
In that case the dimension of $\mathfrak{H}^{(n)}$ equals $12041$ while the dimension of ${\mathfrak{V}}^{(n)}$ is $1141$ (reasonably small). 

On the other hand consider the space $\mathcal{H}^{(n)}\subset\mathrm{End}(\mathbb{C}^2)^{\otimes n}$ of 
$\slt$-invariant and $C$-invariant {(invariant under
simultaneous 
change of sign for all $\sigma_j^a$, $a=1,2,3$)}  operators located on $n$ sites of the spin chain.
We require also that the operators cannot be reduced to
smaller interval, formal definition is given in Section 3.
Let us denote a basis of this space by  $O_a$. The main statement is 
the relations

\begin{align}
O_a\equiv
v_\al X_{\al,a}(n)\,,\label{MAIN}
\end{align}
where $\equiv$ stands for equality of expectation values on a cylinder with
arbitrary Matubara data as we are going to explain. 
This expectation value is denoted by $\langle\cdot\rangle_\mathrm{Md}$.
For $n=10$ the dimension of
$\mathcal{H}^{(n)}$ is $4286$. So, our main problem is to define the matrix $X(n)$ which for $n=10$ is a
{$ 1141\times4286$} matrix.

The Matsubara data consist of {a positive integer $L$,}
the coefficients $\{a_1,\cdots, a_L\}$,  $\{d_1,\cdots, d_L\}$ and
Bethe numbers $\beta_1,\cdots,\beta_m$ ($m\le L/2$), which satisfy the Bethe equations
$$a(\beta_j)Q(\beta_j+1)+d(\beta_j)Q(\beta_j-1)=0,\quad j=1,\cdots, m\,,$$
where
$$a(\la)=\la^L+\sum_{j=1}^La_j\la^{L-j}\,,\ \ d(\la)=\la^L+\sum_{j=1}^Ld_j\la^{L-j}\,,
\ \ Q(\la)=\prod_{j=1}^m({\la}-\beta_j)\,.$$
The matrix $X(n)$ does not depend on the Matsubara data. Hence the main idea:
to take {a set of} simple unphysical Matsubara data
{$\{\mathrm{Md}_j\}$} in order to fix $X(n)$
{through the linear equations
\begin{align}
\langle O_a\rangle_{\mathrm{Md}_j}=X_{a,\al}(n)\langle v_\al\rangle_{\mathrm{Md}_j},
\label{MAINEXP}
\end{align}
}
and then apply it to physically relevant cases. 

We construct the unphysical data as follows. Take the input data
\begin{align}
\mathrm{input}
=\{\beta_1,\cdots,\beta_m, a_{m+1},\cdots, a_{{L}},d_1,\cdots d_{{L}}\},\label{input}
\end{align}
and find the remaining $a_1,\cdots, a_m$ solving the Bethe equations which are
linear for these unknowns. In practice we take the input data as random integers, so,
the procedure is very fast. 

The expectation value {$\langle O_a\rangle_{\mathrm{Md}_j}$}
in the left hand side of \eqref{MAINEXP} is easy to {compute using}  QISM.
In order to compute $\langle v_a\rangle_{\mathrm{Md}_j}$ in the right hand side,
we begin with defining a symmetric function of two variables $\omega(\la,\mu)$
for given Matsubara data \cite{bgks}.

Introduce the kernel and ``half-kernel" functions:
$$K(\la)=\frac  {2}{\la^2-1}\,,\qquad H(\la)=\frac 1{ (\la-1)\la}\,.$$
and the measure
$$d m(\la)=\frac {d\la}{1+\mathfrak{a}(\la)}\,,\qquad \mathfrak{a}(\la)=\frac{a(\la)Q(\la+1)}{d(\la)Q(\la-1)}\,.
$$
We need an auxiliary function defined by the integral equation
$$
G(\eta,\mu)=H(\eta-\mu)+\frac 1 {2\pi i}\oint _\Gamma K(\eta-\sigma)G(\sigma,\mu)dm(\sigma)\,,
$$
where the contour $\Gamma$ goes around the Bethe roots $\beta_1,\cdots,\beta_m$ and the point $\sigma=\mu$. For a finite 
Matsubara chain we have a finite number of Bethe roots for which the equation above reduces to a linear
system for $G(\beta_j,\mu)$. {Then solving it we obtain $G(\beta_j,\mu)$, and $G(\eta,\mu)$ itself as well.}
The function $\omega(\la,\mu)$ is given by 
$$\omega (\la, \mu)=\frac 1 {2\pi i}\oint _{\Gamma '}H(\eta-\la)G(\eta,\mu)dm (\eta)+\frac 1 4 K(\la-\mu)\,,$$
with  $\Gamma'$ containing one more point: $\eta=\la$ .

The Taylor series of $\omega(\la,\mu) $ define an half-infinite matrix $||\omega_{i,j}||$:
\begin{align}
\omega (\la, \mu)=\sum_{i,j=1}^{\infty}\la^{i-1}\mu^{j-1}\omega _{i.j}\,.\label{omcoef}
\end{align}
Then for two {multi-}indices $I$, $J$ of length $l$ define
$$\omega_{I,J}=\det ||\omega_{i_p,j_q}||_{p,q=1,\cdots, l}\,.$$
Then
$$\langle b^*_Ic_J^*\cdot \mathrm{I}\rangle_\mathrm{Md}=\omega_{I,J}\,,$$
consequently
$$\langle v_\al\rangle_\mathrm{Md}=\sum F_{\al,\{I,J\}}\omega_{I,J}\,.$$
We take Matsubara data numerical, so, all these computations are very fast.

\section{Direct computation of expectation values}

First of all we define the Matubara monodromy matrix 
$$T_j(\la)=\begin{pmatrix} A(\la)&B(\la)\\ C(\la)&D(\la)\end{pmatrix}_j\,,$$
where $j$ counts the tensor components in the space direction.

We use the notation $A=A(0)$ {\it etc}. Our goal is to compute
$$\langle O \rangle_\mathrm{Md}=\frac{\langle\Psi|\Tr_{[1,n]}\Bigl(O\ 
T_1\cdots T_n
\Bigr)
|\Psi\rangle}{\langle\Psi|\Psi\rangle}\,,$$
where $|\Psi\rangle$ is the Bethe {vector considered} in the previous section; for
given Matsubara data we write
\begin{align}
&\langle\Psi|=\langle \beta_1,\cdots,\beta_m|=\langle\downarrow|B(\beta_1)\cdots B(\beta_m)
\,.\nn\end{align}
The normalization is provided by the Gaudin formula below. 

{The} Slavnov formula for the scalar product of the Bethe covector $\langle \beta_1,\cdots\beta_m|$ with
an off-shell vector
$| \mu_1,\cdots,\mu_m\rangle=C(\mu_1)\cdots C(\mu_m){\ket{\downarrow}}$, 
{$\mu_j$ being arbitrary is}
\begin{align}
 \langle \beta_1,\cdots {,\beta_m|}\mu_1,\cdots, \mu_m\rangle=\prod\limits_{j=1}^md(\beta_j)d(\mu_j)
\frac{\prod\limits_{i,j=1}^m(\beta_j-\mu_i+1)}{\prod\limits_{i<j}(\mu_i-\mu_j)\prod\limits_{i<j}(\beta_j-\beta_i)}\det({\mathcal N})\,,\label{nik}
\end{align}
where the matrix ${\mathcal N}$ has entries
\begin{align}
N_{i,j}=\frac 1{(\mu_i-\beta_j)}\Bigr(
\frac 1{(\mu_i-\beta_j-1)}
-\frac 1{(\mu_i-\beta_j+1)}\mathfrak{a}(\mu_i)
\Bigr)\,.\nn
\end{align}
{The Slavnov formula \eqref{nik} obviously from the right hand side, is a polynomial in the variables
$\mu_1,\ldots,\mu_m$. In the below $N$ denotes the symmetric polynomial of the variables $\mu_1,\ldots,\mu_m$
 given by the Slavnov formula.}
 
Using the L'Hospital rules { in \eqref{nik}} we get  the Gaudin formula for normalisation. Explicitly, we have
\begin{align}
{\langle \beta_1,\cdots\beta_m|  \beta_1,\cdots\beta_m\rangle}=\prod_{j=1}^ma(\beta_j)
d(\beta_j)\prod_{i\ne j}\frac{\beta_i-\beta_j+1}{\beta_i-\beta_j}\det({\mathcal G})\,,
\end{align}
where the matrix $\mathcal G$ has entries
$$G_{k,l}=\frac{\partial}{\partial\beta_l}\log\mathfrak{a}(\beta_k)\,,\quad k,l=1,\cdots, m\,.$$

We want to compute
$$\langle \beta_1,\cdots\beta_m|X_1X_2\cdots X_N|  \beta_1,\cdots\beta_m\rangle\,,$$
where $X_j$ is one of $A,B,C,D$. {Note that this quantity is zero unless
$\sharp\{j|X_j=B\}=\sharp\{j|X_j=C\}$.}
Starting with the Slavnov formula, we begin the following computation.
Introduce the notations $v(\la)=1/\la,\ u(\la)=v(\la)+1$.

{Let $\bra{\Phi}=\bra{\Psi}X_1X_2\cdots X_N$ for some $X_1,\ldots,X_N$ and set
$$f_{\bra{\Phi}}
(\mu_1,\cdots, \mu_q)=\langle\Phi|\mu_1,\cdots, \mu_q\rangle.$$
Note that this is zero unless $\sharp\{j|X_j=B\}-\sharp\{j|X_j=C\}=q\geq0$. The commutation relation $RTT=TTR$
implies}
\begin{align}
&{f_{\bra{\Phi}A}(\mu_1,\cdots, \mu_q)}
=a(0)\prod_{j=1}^{{q}}u(-\mu_j){f_{\bra\Phi}}(\mu_1,\cdots, \mu_q)\label{A}\\
&-\sum_{j=1}^qa(\mu_j)v(-\mu_j)\prod_{r\ne j}^q  u(\mu_j-\mu_r)
{f_{\bra\Phi}}(\mu_1,\cdots,\widehat{\mu_j},\cdots, \mu_q,0)\,,\nn
\end{align}
\begin{align}
&{f_{\bra{\Phi}D}(\mu_1,\cdots, \mu_q)}
=d(0)\prod_{j=1}^qu(\mu_j){f_{\bra\Phi}}(\mu_1,\cdots, \mu_q)\label{D}\\
&-\sum_{j=1}^qd(\mu_j)v(\mu_j)\prod_{r\ne j}^q  u(\mu_r-\mu_j)
{f_{\bra\Phi}}(\mu_1,\cdots,\widehat{\mu_j},\cdots, \mu_q,0)\,,\nn
\end{align}
\begin{align}
&{f_{\bra{\Phi}B}(\mu_1,\cdots, \mu_q)}
=\sum_{j=1}^q{\Bigl(}a(0)d(\mu_j)v(-\mu_j)\prod_{r\ne j}^qu(-\mu_r)(\mu_r-\mu_j){\label{B}}\\&
+d(0)a(\mu_j)v(\mu_j)\prod_{r\ne j}^q u(\mu_r)(\mu_j-\mu_r)\Bigr)
{f_{\bra\Phi}}(\mu_1,\cdots,\widehat{\mu_j},\cdots, \mu_q)\nn\\
&+\sum_{j>i}\Bigl(d(\mu_i)a(\mu_j)v(-\mu_i)v({\mu_j})u(\mu_j-\mu_i)\prod_{r\ne i,j}^qu(\mu_r-\mu_i)u(\mu_j-\mu_r)\nn\\
&+a(\mu_i)d(\mu_j)v(-\mu_j)v(\mu_i)u(\mu_i-\mu_j)\prod_{r\ne i,j}^qu(\mu_i-\mu_r)u(\mu_r-\mu_j)\Bigr)\nn\\&\times
{f_{\bra\Phi}}({\mu_1,\cdots,\widehat{\mu_i},\cdots,\widehat{\mu_j},\cdots, \mu_q,0})\,.\nn
\end{align}
\begin{align}
&{f_{\bra{\Phi}C}(\mu_1,\cdots, \mu_q)}
={f_{\bra\Phi}}(\mu_1,\cdots,\mu_q,{0})\,.\qquad\qquad\qquad\qquad\qquad{\label{C}}
\end{align}
Notice that the fact that we are doing with $A(0),B(0),C(0),D(0)$ simplifies the general formulae available, for example,
in \cite{BIK}. 

{
{\it Remark}\ 
Let $P_q$ be the space of symmetric polynomials of $q$ variables.
The right hand sides of \eqref{A},\eqref{B},\eqref{C},\eqref{D}
define respectively actions of the operators $A,B,C,D$ on the space $\oplus_{q\geq0}P_q$;
$A$ and $D$ from $P_q$ to itself, $B$ from $P_{q-1}$ to $P_q$ and $C$ $P_{q+1}$ to $P_q$
}

Using the formulae above we compute inductively
$${f_{\bra{\Psi}X_1X_2\cdots X_N}(\mu_1,\cdots, \mu_q)=}
\langle \beta_1,\cdots\beta_m|X_1X_2\cdots X_N|  \mu_1,\cdots\mu_m\rangle\,,$$
and then set $\mu_j=\beta_j$, $j=1,\cdots,m$. However direct application of this procedure to
computer calculation may be very time consuming. Indeed, in order to arrive at symmetric polynomial we have to
factorise the right hand sides of \eqref{A}, \eqref{D}, \eqref{B}. For operators considered in \cite{corr} this is not very hard:
the worst expression we had there is $BTT\cdots TC$ {where $T=A+D$.}

For this expression we have to go up to $m+1$ variables, $m$ is not large (for $n=10$ it suffices to consider $m=2$ at most). That is why the direct procedure works and the
improvement which we explain in what follows only accelerates it. But when computing {the expectation values for all the operators
it is simply impossible to manage for $n=10$ because we have, for example, expressions like}
{
$\underbrace{B\cdots B}_N\underbrace{C\cdots C}_N$, for which the number of variables in the middle
becomes $m+N$.}
Hence the rewriting of the procedure in terms of Schur polynomials
which was mentioned only briefly in \cite{corr} becomes crucial.

%%%%%%%%%%%%%%%%%%%%%%%%%%%%%%%%%%%%%%%%%%%%%%%%%%%%%%%%%%%%%%%%%%%%%%%%%%%%%%%%%%%%%%%%%%%%%%%%%%%%%%%%%%%%%%%%%%%%%%%%
Consider Young diagrams
{$Y_\la$ where $\la=(\la_1,\cdots,\la_n)$, $\la_i\ge\la_{i+1}>0$
is a partition. We set $\#(\la)=n$.}
{It is called the length of $Y_\la$.}
We work in the space $H_q$  whose elements are
$${Y=\sum_{\#(\la)\le q}c_\la Y_\la\,.}$$
{In the below we will identify $Y_\la$ with $\la$. The symbol $\emptyset$ denotes
the empty diagram.}
Define the operation {$\mathrm{cut}_q$ which acts} from $H_{q'}$ with $q'>q$ to $H_{q}$ erasing
all the terms with $\#(\la)>q$. 
Consider the Grassmann space $F_q$ with the 
basis $\psi^*_{k_1}\cdots \psi^*_{k_q}$ ($k_1>\cdots>k_q\ge 0$) .
{We have the usual isomorphism between the spaces $H_q$ and $F_q$.}
\begin{align}
&{\psi^*_{k_1}\psi^*_{k_2}\cdots\psi^*_{k_q}\ \mapsto\ (k_1-(q-1),k_2-(q-2),\cdots ,k_q)_0\,,}\label{maps}\\
&{(\la_1,\cdots,\la_{{n}})\ \mapsto\ \psi^*_{\la_1+q-1}\cdots\psi^*_{\la_{{n}}+q-{{n}}}\psi^*_{q-{{n}}-1}\cdots\psi^*_0}\,{,\ \text{where} \ n\leq q.}\nn
\end{align}
{In the above, $()_0$ }means removing all entries equal to $0$. 
Schur polynomial $s_\la(x_1,\cdots,x_q)$  is the symmetric polynomial
{
$$s_\la(x_1,\cdots,x_q)=\frac{\det ||x_j^{{\la_i+q-i}}||}{\det ||x_j^{q-i}||}\,.$$
The above formula gives an isomorphism between $H_q$ and $P_q$.}

For a given polynomial of one variable $P(x)=\sum_{j=0}^d p_j x^j$ we define
the operator $P\wedge \ F_{q-1}\subset F_q$ multiplying by $\sum_{j=0}^d p_j \psi^*_j$, this operator is
defined {as $P\wedge H_{q-1}\subset H_q$} by the isomorphism \eqref{maps}. We shall also need the simplest Littlewood-Richardson formula for multiplication of a Schur polynomial by
elementary symmetric function {$\s_j$,} which translates as action on $H_q$
\begin{align*}
\sigma_j\circ(\la_1,\cdots, \la _{{n}})
=\sum\limits _J^{\left({{n}}+
\min(j,q-{{n}})\atop j\right)}
\Bigl((\la_1,\cdots, \la _{{n}},\underbrace{0,\cdots,0}_{\min(j,q-{{n}})})+e_J\Bigr)_{\mathrm{order}}\,,
\end{align*}
where $e_J$ are all vectors of dimension ${{n}}+\min(j,q-{{n}})$
with $j$ elements equal to $1$ other elements being $0$, ``order" means that 
we have to drop all the tables in which elements happen to be not ordered, {and}
we also drop all zeros in the final table. 

{The} Slavnov formula {\eqref{nik} gives} the symmetric polynomial $N$ in variables $\mu_1,\cdots,\mu _m$ which
{can} be written as
follows. Define
$$P_j(x)=\frac{x}{x-\beta_j}\(a(x)\frac{Q(x+1)}{x-\beta_j+1}-d(x)\frac{Q(x-1)}{x-\beta_j-1}
\)\,$$
then 
$$N=(-1)^{1/2m(m-1)} \frac{\prod
    d(\beta_j)}{\prod_{i<j}(\beta_i-\beta_j)}
    \cdot P_1\wedge P_2\cdots \wedge P_m\wedge{\emptyset}\in H_m\,.
$$
For us $\beta_1,\cdots ,\beta_m$ are numbers, so, the computation of $N$ is extremely fast.

In what follows we shall need the operation $\mathrm{cut}_q({Y})$
which erases all the Young diagrams in ${Y}$ with lengths greater than $q$.

Now we have to translate the action of the operators $A$, $B$, $C$, $D$.
The operators  $A$, $D$ act from $H_q$ to itself; we have
\begin{align}
A\ {Y}=\mathrm{cut}_q\Bigl(\sum\limits_{k=1}^{q+1}\sigma_{k-1}\circ A_k\wedge {Y}\Bigr)\,,
\quad D\ {Y}=\mathrm{cut}_q\Bigl(\sum\limits_{k=1}^{q+1}\sigma_{k-1}\circ D_k\wedge {Y}\Bigr)\,.\nn
\end{align}
where
$$A_k(x)=(-1)^{k - 1}(x + 1)^{q + 1 - k} a(x)\,,\quad D_k(x)=(-1)^{k - 1}(x - 1)^{q + 1 - k} d(x)\,.$$
The operator $B$ is more complicated. It acts from $H_{q-1}$ to $H_q$. For a polynomial of
two variables {$R(x,y)=\sum_{i,j=0}^dR_{i,j}x^iy^j$ we define the operator $R\wedge H_{q-2}\subset H_q$
mulitiplying by $\sum_{i,j=0}^dR_{i,j}\psi^*_i\psi^*_j$.}
Then
\begin{align}
B\ {Y}
=\mathrm{cut}_q\Bigl(\sigma_q^{-1}\circ\Bigl[\sum_{p=0}^q\sum_{r=0}^{q-1}\sigma_r\circ\sigma_p\circ
\Bigl(P_{p,r}\wedge \mathrm{cut}_{q-1}({Y})+R_{p,r}\wedge\sigma_{q-2}\circ\mathrm{cut}_{q-2}({Y}))
\Bigr)\Bigr]\Bigr)\,,\label{defB}
\end{align} 
where $P_{p,r}$ and $R_{p,r}$ are polynomials of one and two variables, {respectively:}
\begin{align}
&P_{p,r}(x)=(-1)^{r}\sum_{s=r}^{q-1}\[d(0)a(x)  (x + 1)^{q - p}
  x^{s - r}-a(0)d(x) (1-x )^{q - p} x^{s - r} \]\nn\\
  &R_{p,r}(x,y)= d(x) a(y) (-1)^
      { p+r} (y + 1)^{q - 
         p}\sum_{s=r}^{q-1}(x - 1)^{q - 1 - s} y^{s - 
            r}\,.\nn
\end{align}
It can be shown that the expression inside the square brackets in \eqref{defB} consists  of 
Young diagrams of length $q$, not shorter. Hence the $\sigma_q^{-1}$ is applicable: we just
subtract $1$ from all entries of Young diagrams, and drop zeros.

Finally $C$ act from $H_{q+1}$ to $H_q$ simply as
\begin{align}
C\ {Y}=\mathrm{cut}_q({Y})\,,.
\label{defC}
\end{align}

Now we are ready to compute {the} Matsubara expectation value of the right
hand side of \eqref{MAIN}.

Consider
operators  $O$ localised on the interval $[1,n]$. We realise them
{as} linear combinations
of tensor products of $\mathrm{I}, \sigma^\pm, \sigma^3$. We have to take into
account symmetries. First of them is  the translational invariance. The operator
$O$ may contain terms of the form
$$\underbrace{\mathrm{I}\otimes\mathrm{I}}_{k}\otimes\ O^\prime\otimes \underbrace{\mathrm{I}\otimes\mathrm{I}}_{l}\,,$$
with $O^\prime$ localised on $n-k-l$ sites. The expectation value for such operator can be
computed using our procedure for this number of sites. We shall denote by $O^{(n)}_A$ the basis
of operators on $n$ sites irreducible in that way.
It consists of the tensor products which do not contain $\mathrm{I}$ neither at the left nor on the right end. Further, we require $\#(\sigma^+)=\#(\sigma^-)$ to have zero total charge,  $\#(\sigma^3)\equiv 0 (\mathrm{mod}\ 2)$ for
{$C$-invariance.}
Then for {any operator $O$ localised on $n$ sites}
we have the {reduction due to the translational invariance:} 
\begin{align}\mathcal{T}(O)=\{O^{(0)},O^{(2)},\cdots,O^{(n-1)},O^{(n)}\}\,,\label{defT}\end{align}
where $O^{(k)}$ are translationally irreducible {and $C$-invariant} operators on $k$ sites.

First impression is rather discouraging even after
the serious acceleration of the procedure discussed above. 
It has been said that for $n=10$ we are interested in $4286$ $\slt$-invariant,
$C$-invariant  and translationally irreducible operators.
We choose the basis of such operators $O^{(n)}_a$
in certain simplest possible way. 
So, we have a matrix
$$O^{(n)}_a=L_{a,A}^{(n)}O^{(n)}_A\,.$$
The problem is that our procedure does not allow {one} to compute directly for $O^{(n)}_a$, but rather for $O^{(n)}_A$. For $n=10$ the number of the latter 
is  horrifying: $50354$. Fortunately we do not need to compute for all of them
independently. Our computation goes from the left to the right, and, for example
in
$BBBBCCCCBC$ and $BBBBCCCCCB$ the pieces $BBBBCCCC$ coincide,
so, we have to organise the computation in order not to do the same computation twice. This can be done making the total computation reasonably fast.

\section{Summary of computation procedure}

Let us summarise. Consider translationally irreducible operators. First important point is that the equation
\begin{align}
\langle O^{(n)}_a\rangle_\mathrm{Md}\equiv X_{\al,a}(n)\langle v_\al
\rangle_\mathrm{Md}\,,\label{MAIN1}
\end{align}
holds for any Matsubara data {$\mathrm{Md}$}. So, in principle we have an infinite overdetermined
system of equations for the coefficients $X_{a,\al}(n)$. 

Consider our favourite
case $n=10$. We begin with the simplest case $L=1$, $m=0$, and take
$\mathrm{Md}_j$ $(j=1,\ldots,20)$ with 20 random integer input data \eqref{input}. The rank of the matrix
$${
||\langle v_\alpha\rangle_{\mathrm{Md}_j}||_{j,\alpha}=
||F_{\al,\{I,J\}}\omega_{I,J}(\mathrm{Md}_j)||_{j,\alpha,j},}$$
is 15. We can add as many $L=1$, $m=0$ equation as we wish, the rank will not change. So, we proceed to $L=2$, $m=0$ taking 200 equations the rank raises by
almost 200 and {stabilises}. Then we take in addition 300 eqs with $L=3$, $m=0$, 90 eqs with $L=4$, $m=0$, 10 eqs with $L=5$, $m=0$. Adding any other equation with $m=0$ changes nothing, so we proceed to ``one-particle" case taking
10 eqs with $L=2,m=1$,  200 eqs with $L=3,m=1$, 
 325 eqs with $L=4,m=1$,  100 eqs with $L=5,m=1$, more ``one-particle"  equations
 add nothing to rank, and we have to take several ``two-particle" ones
 (fortunately not too many because the computation for them is getting longer).
 We take 10 eqs with $L=4,m=2$, 35 eqs with $L=5,m=2$, 7 eqs with $L=6,m=2$.
 Altogether we have 1307 equations and the rank is 1141. So, we can proceed 
 computing the left hand side of \eqref{MAIN1} for all {these} Matsubara data.

Now we proceed as follows. Construct the matrix
$1307\times 1141$ matrix
$$\mathcal{A}=||F_{\al,\{I,J\}}\omega_{I,J}(\mathrm{Md}_j)||_{j=1,\cdots,1307, \al=1,\cdots 1141}$$
and the $1307\times 4286$ matrix
$$\mathcal{B}=||\langle O^{(n)}_a\rangle_{\mathrm{Md}_j}||_{j=1,\cdots,1307, a=1,\cdots  4286}\,.$$
Put them together
$$||\mathcal{A},\mathcal{B}||\,.$$
By Gaussian procedure which multiplies $GL(1307)$ from the left
we bring the matrix $||\mathcal{A},\mathcal{B}||$ to the form
$$\left|\left|\begin{matrix} I&X(10)\\0&0\end{matrix}\right|\right|\,,$$
where $X(10)$ is the matrix of transformation to the fermionic basis defined above. 
The fact that the  first  $1141$ columns become in this form and
all the rows starting from $1142$-th one vanish is a crucial check
of our entire procedure. It shows that 
the vectors $\langle v_\al\rangle$ are linearly independent, and, more importantly, that
 all the expectation values of our invariant operators are expressible as linear combinations  of $\langle v_\al\rangle$.

 We took some simplest basis of $\slt$-invariant and ${C}$-invariant operators
 $O_a$. {The} price to pay for the simplicity is that we did not {input} the
 orthogonality from the very beginning, and now we have
 to find the operators $\overline{O}^{(n)}_a$ such that
 $${\Tr_{[1,n]}}({O}^{(n)}_a\overline{O}^{(n)}_b)=\delta_{a,b}\,.$$
Introduce $\mathcal{D}(n)_{I,J}$ as
 $$\mathcal{D}(n)_{I,J}(\bullet)={F}_{a,\{I,J\}}(n){\Tr_{[1,n]}}(\overline{O}^{(n)}_a\bullet)\,.$$
 For any Matsubara data we construct $\omega_{i,j}$, and the
 expectation value of any $\slt$-invariant and $C$-invariant  
 translationally irreducible operator $ O^{(n)}$ is
 \begin{align}
 \langle O^{(n)}\rangle=\omega_{I,J}\mathcal{D}(n)_{I,J}( O^{(n)})\,.\label{fin}
 \end{align}
We can drop the requirement of  translational irreducibility applying to any
operator $O$ located on $n$-sites the operator$\mathcal{T}$ \eqref{defT}, and further acting by 
the block-diagonal operators composed of $\mathcal{D}(0)_{I,J},\mathcal{D}(2)_{I,J},\cdots, \mathcal{D}(n)_{I,J}$

 The expressions for $ \mathcal{D}(n)_{I,J}$ become long for $n>6$, so, we cannot present them here, but they are available at \vskip.2cm
 \centerline{\scalebox{.9}{https://www.dropbox.com/sh/l363gixrrgsm95d/AACtqHLdUz7Qj8mD3NSVawMwa?dl=0}}
\vskip.2cm\noindent
 The only Mathematica notebook in this directory gives necessary explanations
 (hopefully sufficient) for application.

If the symmetries are not broken by the Matsubara data  (as it happens for the antiferromagnetic chain at any temperature, but in absence of magnetic field)
we obtain an entire density matrix. 
Let us redo everything in more conventional way. Density matrix $D(n)$ is defined by
$$\langle O\rangle={\Tr_{[1,n]}}(D(n)O)\,,$$
for operators located on $n$ sites. 
In the next section we shall consider the
entanglement entropy {which} is defined by
$$s(n)=-\Tr \(D(n)\log D(n)\)\,.$$
So, in order to compute it we have to diagonalise the density matrix.

We have to take into account the $\slt$-symmetry
of the density matrix. Let {$\epsilon=0\text{ or }1/2$ where $2\epsilon=n\ (\mathrm{mod}\ 2)$.}
We have the orthogonal decomposition
$$V=\(\mathbb{C}^2\)^{\otimes n}=\bigoplus\limits_{j=\epsilon}^{n/2} (M_j\otimes
V_{j})\,,$$
{where $V_j$ is the spin $j$ irreducible representation} of $\slt$ and $ M_j$ is the space of
multiplicities counted by Bratteli diagrams.
The $\slt$-invariant density matrix acts on 
$$M=\bigoplus\limits_{j=\epsilon}^{n/2} M_j\,,$$
but computing the spectrum we have to take into account that the eigenvalues come
with multiplicity $2j+1$. The dimension of $M_j$ equals
{$$\left(n\atop n/2-j\right)-\left(n\atop n/2-j-1\right).$$}
%$C_n^{n/2-j}-C_n^{n/2-j-1}$.
So, for $n=10$ the maximal dimension is that of $M_1$, it is equal to $90$ which
is quite appropriate for the computer diagonalisation. 

The density matrix is obtained from the formulae of the previous section.
We recalculate it in the new basis. Since we are interested in universal formulae,
applicable to any Matsubara data, we compute everything keeping the indices $I,J$ for fermions.

\section{Entanglement entropy at zero temperature}

Now we can proceed to the diagonalisation
of the density matrix. We begin with the antiferromagnetic 
at zero temperature. In that case the function $\omega(\la,\mu)$ is known explicitly:
\begin{align}
&\omega(\la,\mu) =\omega(\la-\mu)\,,\label{om0}
\\& 
\omega(\la)=-\half + 2{\ \log2} + 
 \sum\limits_{k=1}^{\infty}\la^{2 k} \Bigl(2 \zeta(2 k + 1) (1 - 2^{-2 k}) - \half\Bigr)\,.\nn
  \end{align}
With these data we diagonalise the density matrix. The eigenvalues decrease with {the spin} $j$. The most striking example is given by $j=n/2$, corresponding block is $1\times 1$, it
coincides with the vacuum formation probability. 
The numerical values:
\begin{align}
&P(2)=0.102284273146684897,\nn\\&  P(3)=0.00762415812490254761,\nn\\&
P(4)=0.000206270046519527063,\nn\\& P(5)=2.01172595898884905\cdot10^{-6},\nn\\& 
 P(6)=7.06812753309203896\cdot10^{-9},\nn\\& P(7)=8.93090684226941650\cdot10^{-12},\nn\\& 
 P(8)=4.05749505255338289\cdot10^{-15},\nn\\&P(9)= 6.62359212493539014\cdot10^{-19},\nn\\& 
 P(10)=3.88481154904260358\cdot10^{-23}\,,\nn
\end{align}
are in very good agreement with asymptotics \cite{J-lukyanov} which looks as follows:
$$P(n)\simeq An^{-\frac 1 {12}}\(\frac{\Gamma^2(1/4)}{\pi\sqrt{2\pi}}\)^{-n^2}\quad(n\rightarrow\infty)\,.$$
The constant $A$ is unknown, in \cite{J-lukyanov}
it is estimated as $A=0.841$. Our
data show that 
$$X(n)=\log P(n)+\log\(\frac{\Gamma^2(1/4)}{\pi\sqrt{2\pi}}\)n^2+\frac 1 {12}\log n\,,$$
slightly oscillates around $\sim \log(0.841)$. So, we ask a question whether the next correction to the
asymptotics is purely oscillating or there is a non-oscillating part. To answer this question we 
compute:
\begin{align}&\exp\Bigl(\frac 1 4(X(10)+2X(9)+X(8))\Bigr)=0.8412645021372811\,,\nn\\
&\exp\Bigl(\frac 1 4(X(9)+2X(8)+X(7))\Bigr)=0.8412642481617325\,.\nn
\end{align}
This computation convinces us that the power corrections are purely oscillating, and that the good approximation for $A$
is
$$A=0.841264(5)\,.$$

In the Appendix we give the eigenvalues of the density matrix with 11 digits accuracy
(with this accuracy the eigenvalues disappear for high spins).

Here are  entanglement entropies with $24$ digits
\begin{align}
&s(2)= 0.95367162656978945738557\nn\\&s(3)= 1.09690078367655639608404\nn\\&s(4)= 
  1.19547447383418925567332\nn\\&s(5)= 1.27102739309231825158036\nn\\&s(6)= 
  1.33247760568637557112695\nn\\&s(7)= 1.38430489902101253089084\nn\\&s(8)= 
  1.42913854287157243504956\nn\\&s(9)= 1.46864496929391162170464\nn\\&s(10)= 
  1.50396085818734543200735\nn\,.
  \end{align}
To verify that the first five numbers agree with those of \cite{takahashi} one has to 
pass form natural logarithms to binary ones.

The CFT predicts \cite{cft} that
{
$$s(n)\simeq \frac 1 3 \log n+C\,.$$
}
The following figure shows that we are rather close to the conformal limit  
\vskip .4cm
\centerline{\includegraphics[height=7cm]{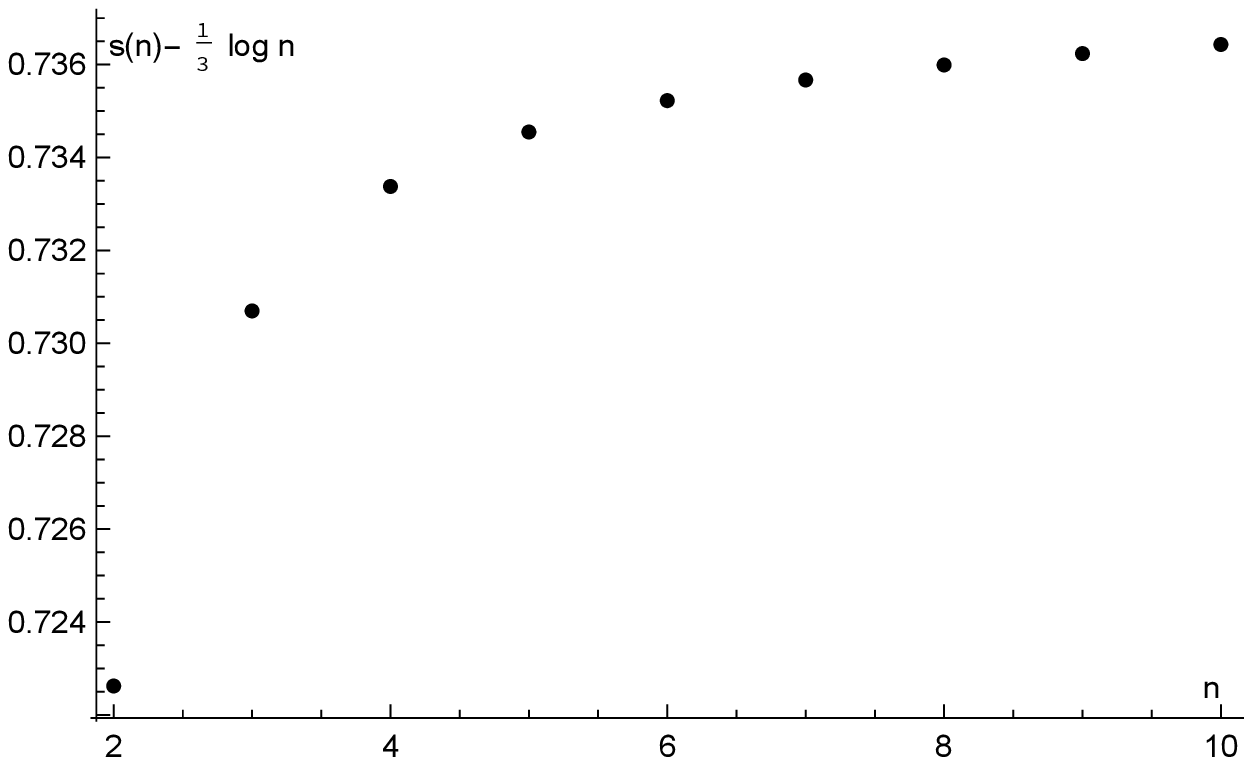}}
\vskip .5cm

\section{Computation of $\omega(\la,\mu)$ with temperature}

At finite temperature the function $\omega(\la,\mu)$ can be computed
only numerically. For temperature equal to $T$ we shall denote it by
$\omega_T(\la,\mu)$, in particular, the function \eqref{om0} will be denoted by
$\omega_0(\la,\mu)$ from now on.
{We compute 10 Taylor coefficients in each variable of the function
$\omega_T(\la,\mu)$.}
The {main} problem here is that we need to know it with very high precision: our answers contain the sign changing sums 
with huge rational coefficients
of determinants made of the Taylor coefficients of $\omega$ with sizes up to $5\times 5$. So, we have to find a good and controllable way of computation.

We follow the definitions of the Section \ref{fermionic}, but now our goal is different:
we are interested in quite special Matsubara data, staggering inhomogeneities and limit $L\to\infty$. 
It will be convenient to change the variables to  $\la=ix$, $\mu=iy$, {\it etc}. 
We do not go into details which are well-known from \cite{klumper}
presenting directly the equation for $\mathfrak{a}(x)$:
\begin{align}
\log\mathfrak{a}(x)=\frac{h_0(x)}T+\int\limits_{C}K(x-y)\log\(1+\frak{a}(y)\)dy\,,
\qquad {h_0(x)}=\frac 1 {x(x+i)}\,,\label{eqa}
\end{align}
which holds for the ground state which is of interest to us.
We slightly change the definition of $K$:
$$K(x)=-\frac 1 {\pi (x^2+1)}\,.$$
The contour $C$ goes around the Bethe roots, {and}
the change of variables was performed in order to make them real. 
The Bethe roots are situated symmetrically with respect to the point {$x=0$, and accumulate}
at the point $x=0$ where the function $\mathfrak{a}(x)$ has essential singularity. 
The maximal Bethe root, $\beta_\mathrm{max}$ grows logarithmically with $1/T$. We shall take $C$ as ellipse 
\begin{align}x(\phi)=-R\cos(\phi)-it\sin\phi, \quad 0\le \phi<2\pi\,.\label{ell}\end{align}
We shall denote by $C_-$ the part of $C$ situated in lower half plane,
and by $C_+$ the part of $C$ in the upper half plane with reversed orientation. 
The parameter $R$ must be bigger {than}
$\beta_\mathrm{max}$ while $0<t<1$. We shall take $t=2/5$. For $R$ there is a simple check: 
for given $T$ solve the equation and make sure
that $\log\mathfrak{a}(R)/i<{\pi}$. 

In order to make the iterative procedure for \eqref{eqa} efficient we use Destri-DeVega trick. By Schwarz principle
$$\mathfrak{a}(x)=\frac 1 {\overline{\mathfrak{a}(\bar x)}}$$
For sufficiently small temperatures $|\mathfrak{a}(x)|<1$ {holds}
for $x\in \mathbb{C}^+$, and {it gets very small when $x$ is close to $0$,}
which is the point of essential singularity.
So, we rewrite \eqref{eqa} as
\begin{align}
\log\mathfrak{a}(x)=\frac {h(x)} T -\int\limits_{C_+}R(x-y)\log\( 1+\mathfrak{a}(y)\)dy+
\int\limits_{C_-}R(x-y)\log\( 1+\overline{\mathfrak{a}(y)}\)dy\,,\label{neweqa}
\end{align}
where {$R(x)$} is the resolvent of the operator $I-K$ on the interval $[-R,R]$,
$$h=(I+R)h_0\,.$$

So, our first task is to solve with good precision the equation
\begin{align}R(x,y)=K(x-y)+\int\limits _{-R}^RK(x-z)R(z,y)dz\,.\label{eqR}
\end{align}
Simple experiments show that in order to go to temperatures as low as $1/200$ we need $R=2$. Then for temperatures higher than $1/10$ we can switch to $R=1$. 
These are two cases which we shall consider. 
The main problem here is {at the ends of integration; simple-minded discretisation}
gives very bad results for finite intervals. In order to avoid this problem we use the double exponential method \cite{MM}. 
To integrate a function $f(x)$ from $-R$ to $R$
we introduce
{\begin{align} g(t)=-1+\frac 4 \pi \arctan(\exp(c\sinh(t)))\,,\label{gfunction}\end{align}}
and 
use the approximation
$$\int_{-R}^{R}f(x)dx\simeq hR\sum_{k=-N}^{N}f(Rg(hk))g'(hk)\,.$$
{Actually, the function \eqref{gfunction} is different from
the ones used traditionally, it was introduced rather recently \cite{ooura}. It makes
the numerical integration procedure rather fast and marvellously precise.}
We always take the parameter $c$ equal to $1/10$.
For the rest of parameters we take
{
\begin{align}
&h=1/20,\ \ N=200,\quad \mathrm{for}\ \ R=1\,,\nn\\
&h=1/25,\ \ N=250,\quad \mathrm{for}\ \ R=2\,.\nn
\end{align}
}
This gives astonishingly good precision of 70 digits for the functions of the type of $K(x)$.

Then we continue the resolvent to $C_\pm$ by virtue of the equation \eqref{eqR} and its
transposition (the operators are self-adjoint), and apply the same double exponential
trick for the integrals over $\phi$ in \eqref{neweqa} with the parametrisation
\eqref{ell}. For these integrals we shall
use other parameters (mostly for computations to {follow, which} need higher
precision):
{
\begin{align}
&h=1/30,\ \ N=300,\quad\mathrm{for}\ \ R=1\nn\\
&h=1/40,\ \ N=400,\quad \mathrm{for}\ \ R=2\,.\nn
\end{align}
}

We begin with equations for {$\omega(x,y,T)=\omega_T(ix,iy)$,} which is the result of 
a {procedure, similar}
to that we used to modify the equation for $\log\mathfrak{a}$ \cite{HGSIV}.
We have
$$\omega(x,y,T)={\omega_1(x,y)}+\omega_2(x,y,T)\,.$$
The first term does not depend on  temperature, but it depends on $R$ which
has {to be chosen} for a given range of temperatures as has been explained.
We have
\begin{align}
\omega_1(x,y)=\frac 1 {2\pi }\int\limits_{C_-}f(z-x)F(z,y)dz+\frac{\pi}2K(x-y)\,,
\end{align}
where
\begin{align}
F(x,y)=f(x-y)+\int\limits_{C_-}R(x,z)f(z-y)dz,\qquad f(x)=\frac{i}{x(x+i)}\,,\nn
\end{align}
For $\omega_2(x,y,T)$ we have
\begin{align}
\omega_2(x,y,T)=\frac 1 \pi\Bigl(\int\limits_{C_+}F(z,x)G(z,y)d\overline{m}(z)
+\int\limits_{C_-}F(z,x)G(z,y)dm(z)\Bigr)
\,,\label{om2}
\end{align}
where  the measure is as before
\begin{align}
&dm(x)=\frac {dx} {1+ \frak{a}(x)}\,,\nn
\end{align}
and the auxiliary function satisfying the equation
\begin{align}
&G(x,y)=F(x,y)-\int\limits_{C_+}R(x-z)G(x,y)d\overline{m}(z)-\int\limits_{C_-}R(x-z)G(x,y)dm(z)
\end{align}

We  need not the  function $\omega(x,y,T)$, but rather its Taylor coefficients
{
$$\omega(x,y,T)=\sum_{j,k=1}^\infty\omega_{j,k}x^{j-1}y^{k-1}\,,$$
for $1\leq j,k\leq10$.}
To get them we begin with the functions {$f_k(x)$} which are Taylor coefficients of 
$f(x-y)$ in $y$, and define {$F_k(x)$, $G_k(x)$} in obvious way. Then we plug 
{$f_k(x)$, $F_k(x)$, $G_k(x)$} into the definitions of $\omega_1$, $\omega_2$ getting
directly {$\omega_{j,k}(T)$}. The trouble here is that the functions {$f_k(x)$, and consequently $F_k(x)$, $G_k(x)$, have poles of order $k-1$} at $x=0,-i$.
These poles are close to the integration contour which makes the integrands rather sharp.
Numerical integration of such functions needs too much of precision. This concerns
especially the function $\omega_1(x,y)$ where the singularities {coming doubly} from
two multipliers. It is not so bad, but still unpleasant for $F(x,y)$. Finally,
for small enough temperature this problem does not concern $G$ and
$\omega_2$: the measure  $dm(x)$is very small near $\mathrm{Re}(x)=0$, so, the contribution
of singularities is dumped by it. Let us explain how to treat this problem for $F$ and
$\omega_1$.

Fortunately, we have an explicit solutions for $T=0$. The {corresponding}
function $F_0(x,y)$ satisfying
$$F_0(x-y)-\int\limits _{-\infty-i0}^{\infty-i0}K(x-z)F_0(z-y)=f(x-y)\,,$$
is simply
$$F_0(x)=\frac{\pi}{\sinh(\pi x)}\,.$$
Certainly the singularitie at $x=y$ cancel in
$$\Delta F(x,y)=F(x,y)-F_0(x-y)\,.$$
For this function one immediately derives
\begin{align}
&\Delta F(x,y)=d(x,y)+\int\limits_{-R}^RR(x,z)d(z,y)dz\,,\label{neweqF}\\ &
 d(x,y)=-\Bigl(\int\limits_{-\infty}^{-R}+\int\limits_{R}^{\infty}\Bigr)K(x-z)F_0(z-y)dz\,.\nn
\end{align}
Now we rewrite the definition of $\omega_1$:
\begin{align}
\omega_1(x,y)=\omega(i(x-y))-\frac 1 {2\pi}\Bigl(\int\limits_{-\infty}^{-R}+\int\limits_{R}^{\infty}\Bigr)
f(z-x)F_0(z-y)dz+\frac 1 {2\pi}\int\limits_{C_-}f(z-x)\Delta F(z,y)dz\,,\label{newom1}
\end{align}
where $\omega(\la)$ is defined in \eqref{om0}.
In the last integral {singularities close to the contour of integration} remain
in $f(z-x)$, but {they} do not double with the singularities of $F(z,y)$, and
we can arrive at good precision. 

Let us summarise our procedure. For given $R$ (we take $R=1,2$) we first solve
{the equation for $R$ \eqref{eqR} by iterations, with great precision ($60$ digits).}
Then we find $R(x,y)$ {with $x,y\in C_\pm$} using the equation {\eqref{eqR}}.
Then we find $\Delta F$ from
{\eqref{neweqF} }
and $\omega_1$ from \eqref{newom1}. Now we start to work with temperature.
{First we solve the equation for $\log \mathfrak{a}$ \eqref{eqa}}
and verify that $1/i\log\mathfrak{a}(R)<\pi$. Now we solve by iterations {the} equation for $F$ \eqref{neweqF}, finally we
find $\omega_2$ \eqref{om2}.

Let us mention checks which we have performed. Our numerical integration over the
real line and the ellipse for the resolvent can be checked by the Cauchy theorem:
$$R(x,y)=K(x,y)-\int\limits_{C_-}K(x,z)R(z,y)dz\,, \quad etc.$$
More crucial is to check that $\omega_1(T)_{i,j}$ vanish for $i+j$ odd. This is really
nontrivial when we apply \eqref{newom1}, and if the precision is lost somewhere
it is immediately felt. Finally, we take $R=2$ starting from $T=1/200$ than from $T=1/10$ we can switch to $R=1$ which is more economic for the computer time.
The check is to see that for $T=1/10$ both $R=1$ and $R=2$ give the same result. 

\section{Entanglement entropy at finite temperature}

We shall consider temperatures from $1/200$ to $3/8$. With temperature the fates of different
expectation values differ. For example, the correlation function $-\langle\sigma^3_1\sigma^3_{10}\rangle$ obviously decays while the vacuum formation probability
grows: it is more probable to find a piece of ferromagnetic chain when
the antiferromagnetic order is destroyed by temperatures. This is illustrated on the figures {below:}
\vskip .4cm
\includegraphics[height=4.5cm]{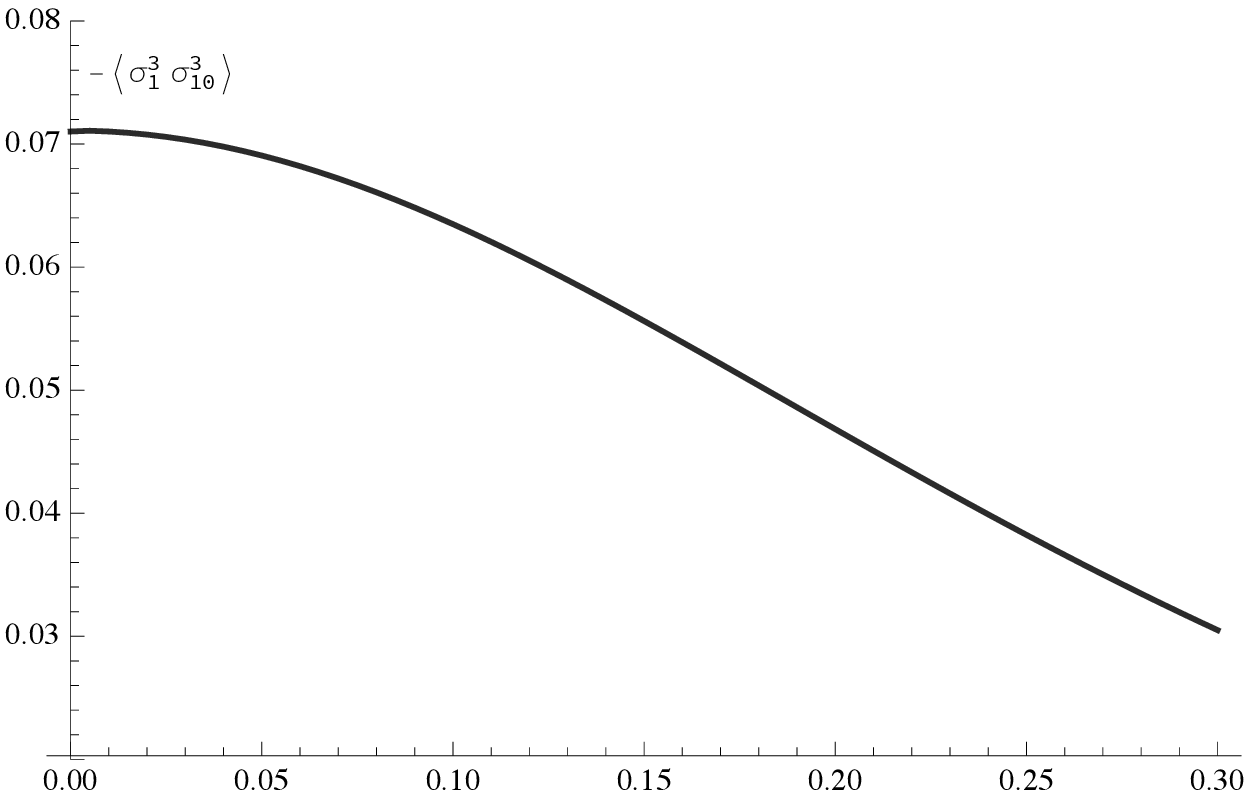}\hskip.7cm\includegraphics[height=4.5cm]{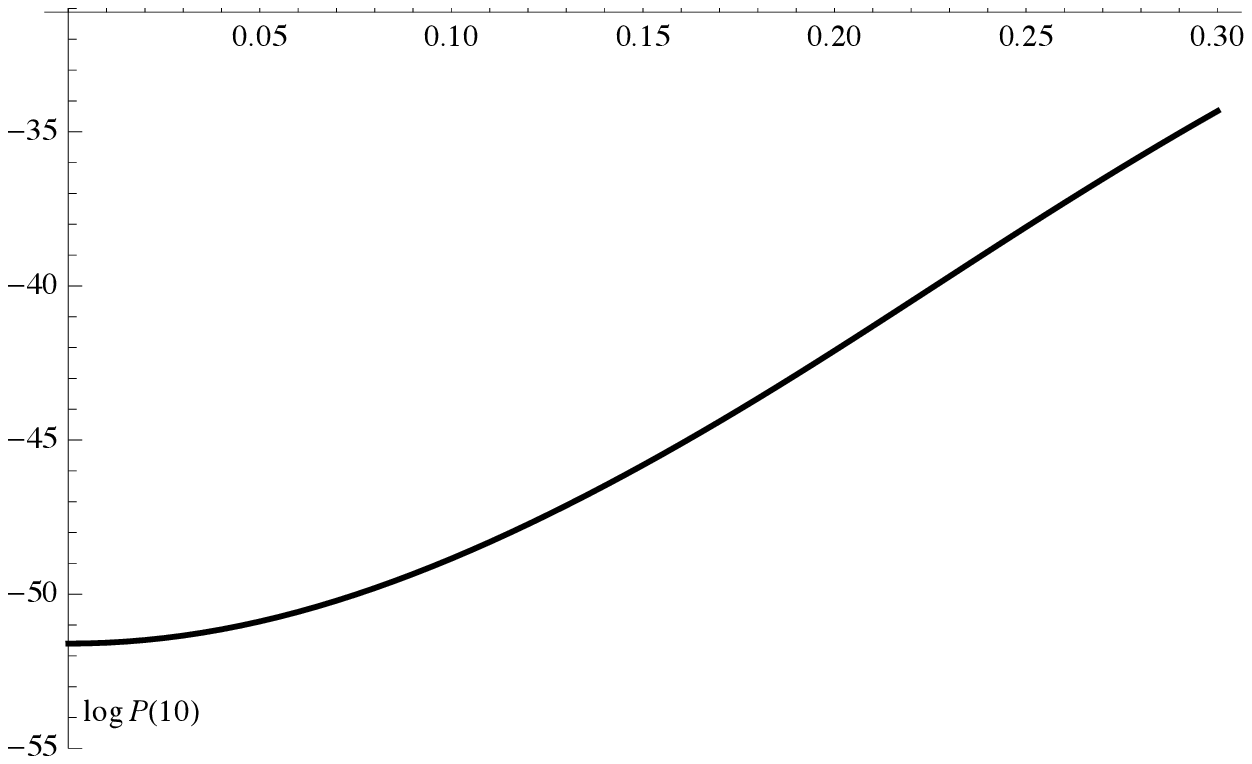}
\vskip .5cm

We compute the entanglement entropy $s(n,T)$ for low temperatures
(up to $T=3/8$). We shall be interested in the difference $s(n,T)-s(n,0)$ for which  we
want to verify two things. First, is it true that for $n=8,9,10$ we are approaching
the scaling limit which means that the difference in question becomes
a function of $nT$? Second, is it true that we are not far from the CFT 
\cite{korepin,cardy} which predicts
$$s(n,T)-s(n,0)\simeq \frac 1 3\log\(\frac{\sinh(nT)}{nT}\)\,.$$
In the right hand side we took into account all necessary normalisations. 
The answers to both question are in the following table in which $s(n,T)-s(n,0)$
are given for $n=8,9,10$ with $nT$ varying from $.05$ to $2$ with step $.05$.
We see that the values of  $s(n,T)-s(n,0)$ are close for $n=8,9,10$, and the last one
is reasonably close to the CFT prediction. Certainly, the difference grows for 
large $nT$. It is interesting to notice that for small $nT$ the values of $s(n,T)-s(n,0)$ are lower than the CFT prediction, around $nT=1.2$ they cross the CFT prediction, and start to be a little larger. 

\newpage
\scalebox{1}{\begin{tabular}{|r|r|r|r|r|}
  \hline
  $nT$\ \ \  &$n=8\ \quad$ & $n=9\ \quad$  & $n=10\ \quad$ &$\mathrm{CFT}\qquad$\\
  \hline 
0.05& 0.00013718326& 0.00013758335& 0.00013786523& 
  0.00013887732\\ \hline0.10& 0.00054888502& 0.00055046054& 0.00055156807& 
  0.00055537049\\ \hline0.15& 0.00123508732& 0.00123856920& 0.00124101147& 
  0.00124906384\\ \hline0.20& 0.00219552716& 0.002201589066& 0.00220583162&
   0.00221926676\\ \hline0.25& 0.00342971960& 0.00343896461& 0.00344541993&
   0.00346501700\\ \hline0.30& 0.00493696730& 0.00494991449& 0.00495893249&
   0.00498508514\\ \hline0.35& 0.006716367454& 0.00673343866& 
  0.006745297245& 0.00677798038\\ \hline0.40& 0.008766818592& 
  0.00878832527& 0.00880322110& 0.00884195738\\ \hline0.45& 0.01108702763& 
  0.01111315797& 0.01113119736& 0.011175024250\\ \hline0.50& 0.01367551758&
   0.01370632378& 0.01372751381& 0.013774951538\\ \hline0.55& 
  0.01653063593& 0.01656602169& 0.01659026149& 0.016639282089\\ \hline0.60&
   0.01965056377& 0.01969027193& 0.01971734418& 
  0.019765341782\\ \hline0.65& 0.023033325512& 0.02307692586& 
  0.02310648849& 0.023150250959\\ \hline0.70& 0.02667679930& 
  0.026723676529& 0.02675525456& 0.026790936485\\ \hline0.75& 
  0.030578728022& 0.03062806972& 0.030661047258& 
  0.030684144322\\ \hline0.80& 0.03473673076& 0.03478751545& 0.03482112784&
   0.034826452504\\ \hline0.85& 0.039148314771& 0.03919929988& 
  0.03923262593& 0.039214284423\\ \hline0.90& 0.04381088783& 0.04386059748&
   0.04389255176& 0.043843922307\\ \hline0.95& 0.04872177091& 
  0.04876848353& 0.04879780864& 0.048711520796\\ \hline1.00& 0.05387821125&
   0.05391994665& 0.05394520545& 0.053813120524\\ \hline1.05& 
  0.05927739556& 0.05931190153& 0.05933146930& 0.059144661603\\ \hline1.10&
   0.064916463659& 0.06494120167& 0.064953257954& 
  0.064701996941\\ \hline1.15& 0.07079252225& 0.07080465212& 0.07080717234&
   0.07048090530\\ \hline1.20& 0.07690265901& 0.076899022173& 
  0.07688976878& 0.07647710404\\ \hline1.25& 0.08324395685& 0.08322105804& 
  0.08319757106& 0.08268626146\\ \hline1.30& 0.08981350845& 0.08976749539& 
  0.089727082245& 0.08910400872\\ \hline1.35& 0.09660843075& 0.09653507178&
   0.09647479626& 0.09572595123\\ \hline1.40& 0.10362587957& 
  0.103520539027& 0.10343720911& 0.10254767960\\ \hline1.45& 
  0.110863063859& 0.11072067531& 0.11061082984& 0.10956477991\\ \hline1.50&
   0.11831725958& 0.11813229713& 0.11799219117& 0.11677284346\\ \hline1.55&
   0.12598582287& 0.125752270925& 0.12557785975& 
  0.12416747595\\ \hline1.60& 0.13386620216& 0.13357752430& 0.13336444609& 
  0.13174430595\\ \hline1.65& 0.14195594901& 0.14160505680& 0.14134861406& 
  0.13949899283\\ \hline1.70& 0.15025272729& 0.149831949990& 
  0.149527090012& 0.14742723403\\ \hline1.75& 0.15875432052& 0.15825537676&
   0.15789667142& 0.15552477175\\ \hline1.80& 0.16745863710& 0.16687260977&
   0.16645423495& 0.16378739896\\ \hline1.85& 0.17636371328& 0.17568102870&
   0.17519674406& 0.17221096492\\ \hline1.90& 0.18546771377& 0.18467812640&
   0.184121255818& 0.18079138004\\ \hline1.95& 0.19476892996& 
  0.193861513653& 0.19322492710& 0.18952462022\\ \hline2.00& 
  0.204265775830& 0.20322892251& 0.20250501998& 0.19840673068\\ \hline
   \end{tabular}}
\newpage
  
 For better visualisation we compare the $n=10$ results (dashed line) with  the
 CFT curve up to $nT=3$. We observe a reasonable agreement. 
 \vskip .2cm
\centerline{\includegraphics[height=8cm]{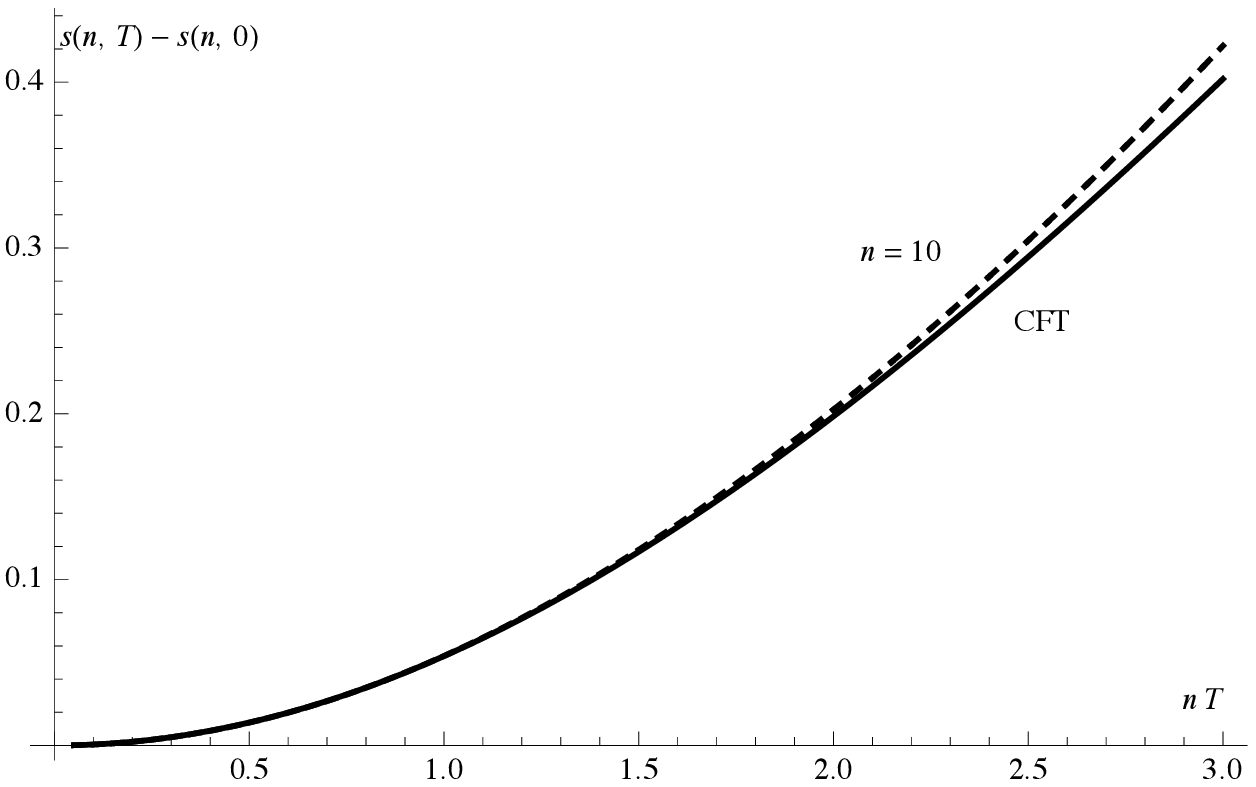}}
\vskip .5cm
\vskip .2cm

{\bf Acknowledgements}  TM thanks to T. Ooura for suggesting his kernel function used in our calculation.

\section{Appendix}
In this appendix we give the eigenvalues of the density matrix for $T=0$
with accuracy $10^{-11}$. For high spins the eigenvalues become too small, {and therefore}
we do not write them. \\

{$n=2,\ j=0,1$}
\begin{align}
&\{0.69314718056\}, \nn\\&\{0.10228427315\}\,.\nn
\end{align}

$n=3,\ j=1/2,3/2$
\begin{align}
&\{0.450771338685, 0.03398034507\}, \nn\\&\{0.007624158125\}\,.\nn
\end{align}

$n=4,\ j=0,1,2$
\begin{align}
&\{0.61451589297, 0.00365561121\},\nn\\& \{0.12071380424, 0.00552473720, 
0.00069384043\}, \nn\\&\{0.000206270047\}\,.\nn
\end{align}

$n=5,\ j=1/2,3/2,5/2$
\begin{align}
&\{0.42478947699, 0.04837782416, 0.00132787973, 0.00016215953, 
  0.00002079330\}, \nn\\
&\{0.01220782094, 0.00041374155, 0.00003079567, 
 {5.55739\cdot10^{-6}}\}, \nn\\
&{\{2.01173\cdot10^{-6}}\}\,.\nn
\end{align}
\newpage

$n=6,\ j=0,1,2,3$
\begin{align}
&\{0.57225072096, 0.00689732739, 0.00012390859, 0.00001153518, 2.1124\cdot10^{-7}\},\nn\\
& \{0.12810808044, 0.00963410772, 0.00146363784, 0.00020810707, 0.00003475259,\nn\\
& 2.69435\cdot10^{-6}, 1.59341\cdot10^{-6}, 2.7386\cdot10^{-7}, 5.023\cdot10^{-8}\},\nn\\
& \{0.00045834467, 0.00001216336, 6.7394\cdot10^{-7}, 7.206\cdot10^{-8}, 1.690\cdot10^{-8}\}, \nn\\
&\{7.07\cdot10^{-9}\}\,.\nn
\end{align}

$n=7,\ j=1/2,3/2,5/2$
  \begin{align}
  &\{0.40741354415, 0.05661439956, 0.00274447210, 0.00041094696, 
  0.00006055511,\nn\\& 0.00004663152, 5.80502\cdot10^{-6}, 1.09218\cdot10^{-6}, 
  3.7937\cdot10^{-7}, 6.181\cdot10^{-8}, 3.019\cdot10^{-8}, \nn\\&4.47\cdot10^{-9}, 8.4\cdot10^{-10}, 
  1.6\cdot10^{-10}\},,\nn\\& \{0.01533056579, 0.00089067320, 0.00008573919, 
  0.00001697604,0.00001524263, \nn\\& 1.72604\cdot10^{-6}, 3.4884\cdot10^{-7}, 
  6.306\cdot10^{-8}, 1.710\cdot10^{-8}, 1.208\cdot10^{-8}, 1.46\cdot10^{-9}, \nn\\& 9.1\cdot10^{-10}, 
  2.0\cdot10^{-10}, 5.\cdot10^{-11}\}, \nn\\& \{6.30299\cdot10^{-6}, 1.3831\cdot10^{-7}, 5.91\cdot10^{-9}, 
  4.8\cdot10^{-10}, 7.\cdot10^{-11}, 2.\cdot10^{-11}\}\,.\nn
  \end{align}
  
$n=8,\ j=0,1,2,3$
  \begin{align}
  &\{0.54407108951, 0.009518029040, 0.00031430987, 0.00003722014, 
  4.34242\cdot10^{-6},\nn\\& 8.6837\cdot10^{-7}, 4.4767\cdot10^{-7}, 2.109\cdot10^{-8}, 1.263\cdot10^{-8},
   5.6\cdot10^{-10}, 2.0\cdot10^{-10}, 3.\cdot10^{-11}\} ,\nn\\&\{0.13192740945, 
  0.01273100394, 0.00217334341, 0.00049770442, 0.00009211769 ,\nn\\&
  9.33699\cdot10^{-6}, 7.06799\cdot10^{-6}, 5.77728\cdot10^{-6}, 1.29977\cdot10^{-6}, 
  1.10141\cdot10^{-6}, 2.1627\cdot10^{-7}, \nn\\&1.1066\cdot10^{-7}, 9.385\cdot10^{-8}, 1.705\cdot10^{-8},
   5.46\cdot10^{-9}, 3.62\cdot10^{-9}, 2.72\cdot10^{-9}, 6.6\cdot10^{-10}, \nn\\&3.4\cdot10^{-10}, 
  1.4\cdot10^{-10}, 7.\cdot10^{-11}, 4.\cdot10^{-11}\},\nn\\& \{0.00070629696, 0.00003306502, 2.45652\cdot10^{-6}, 
  4.7394\cdot10^{-7}, 3.0235\cdot10^{-7}, 7.450\cdot10^{-8},\nn\\& 4.116\cdot10^{-8}, 5.43\cdot10^{-9}, 
  1.33\cdot10^{-9}, 1.19\cdot10^{-9}, 1.8\cdot10^{-10}, 5.\cdot10^{-11}, 3.\cdot10^{-11}, 1.\cdot10^{-11}\},\nn\\& \{3.159\cdot10^{-8}, 5.9\cdot10^{-10}, 2.\cdot10^{-11}\}\,.\nn
   \end{align}
\newpage
 $n=9,\ j=1/2,3/2,5/2,7/2$
  \begin{align}
  &\{0.39446858225, 0.06203960539, 0.00404495595, 0.00068660207, 
  0.00012681903,\nn\\& 0.00011130269, 0.00001809518, 3.66388\cdot10^{-6}, 
  1.57011\cdot10^{-6}, 1.53455\cdot10^{-6},\nn\\& 2.8038\cdot10^{-7}, 2.1435\cdot10^{-7}, 
  1.4717\cdot10^{-7}, 4.352\cdot10^{-8}, 2.421\cdot10^{-8}, 1.612\cdot10^{-8}, \nn\\&4.91\cdot10^{-9}, 
  2.99\cdot10^{-9}, 1.98\cdot10^{-9}, 9.6\cdot10^{-10}, 6.1\cdot10^{-10}, 3.2\cdot10^{-10}, 
  1.1\cdot10^{-10},\nn\\& 6.\cdot10^{-11}, 5.\cdot10^{-11}, 1.\cdot10^{-11}, 0.\cdot10^{-11}\},\nn\\& \{0.01764053114, 0.00135269525, 0.00015298092, 
  0.00004270988, 0.00003235274, \nn\\&5.57512\cdot10^{-6}, 1.20176\cdot10^{-6}, 
  5.2612\cdot10^{-7}, 2.8393\cdot10^{-7}, 8.094\cdot10^{-8}, 6.731\cdot10^{-8},\nn\\& 5.960\cdot10^{-8}, 
  1.449\cdot10^{-8}, 8.14\cdot10^{-9}, 5.22\cdot10^{-9}, 4.09\cdot10^{-9}, 1.23\cdot10^{-9}, 
  9.8\cdot10^{-10}, \nn\\&7.0\cdot10^{-10}, 3.0\cdot10^{-10}, 1.1\cdot10^{-10}, 1.\cdot10^{-10}, 
  8.\cdot10^{-11}, 3.\cdot10^{-11}, 2.\cdot10^{-11}, 2.\cdot10^{-11}\},\nn\\& \{0.00001225502, 4.8365\cdot10^{-7}, 2.849\cdot10^{-8}, 
  5.91\cdot10^{-9}, 2.72\cdot10^{-9}, 4.2\cdot10^{-10},\nn\\& 4.2\cdot10^{-10}, 1.2\cdot10^{-10}, 
  4.\cdot10^{-11}, 0.\cdot10^{-11}\},\nn\\& \{6.\cdot10^{-11}\}\,.\nn
  \end{align}
  
  $n=10,\ j=0,1,2,3$
  \begin{align}
  &\{0.52322247016, 0.01165676559, 0.00053353501, 0.00007341532, 
  0.00001374860,\nn\\& 2.03723\cdot10^{-6}, 1.66763\cdot10^{-6}, 1.4508\cdot10^{-7}, 
  1.0705\cdot10^{-7}, 5.471\cdot10^{-8}, 1.737\cdot10^{-8},\nn\\& 3.36\cdot10^{-9}, 1.36\cdot10^{-9}, 
  9.6\cdot10^{-10}, 5.5\cdot10^{-10}, 2.2\cdot10^{-10}, 4.\cdot10^{-11}, 3.\cdot10^{-11},\nn\\& 3.\cdot10^{-11},
   2.\cdot10^{-11}\},\nn\\& \{0.13415188237, 0.01516080455, 
  0.00280724281, 0.00081234228, 0.00016161568,\nn\\& 0.00002156505, 
  0.00001938501, 0.00001238378, 4.24831\cdot10^{-6}, 2.54338\cdot10^{-6}, \nn\\&
  5.2611\cdot10^{-7}, 4.3033\cdot10^{-7}, 3.7137\cdot10^{-7}, 2.3056\cdot10^{-7}, 7.255\cdot10^{-8},
   4.556\cdot10^{-8},\nn\\& 2.913\cdot10^{-8}, 1.613\cdot10^{-8}, 1.522\cdot10^{-8}, 4.47\cdot10^{-9}, 
  3.92\cdot10^{-9}, 3.90\cdot10^{-9}, 2.20\cdot10^{-9},\nn\\& 8.6\cdot10^{-10}, 7.5\cdot10^{-10}, 
  5.1\cdot10^{-10}, 2.8\cdot10^{-10}, 2.7\cdot10^{-10}, 1.9\cdot10^{-10}, 1.7\cdot10^{-10}, 
  7.\cdot10^{-11},\nn\\& 4.\cdot10^{-11}, 3.\cdot10^{-11}, 2.\cdot10^{-11}, 1.\cdot10^{-11}\},\nn\\& \{0.000938440865, 0.00005918266, 5.29807\cdot10^{-6}, 
  1.58240\cdot10^{-6}, 7.2140\cdot10^{-7},\nn\\& 1.8468\cdot10^{-7}, 1.6292\cdot10^{-7}, 
  2.367\cdot10^{-8}, 1.706\cdot10^{-8}, 6.67\cdot10^{-9}, 6.05\cdot10^{-9}, 1.73\cdot10^{-9}, \nn\\&
  1.09\cdot10^{-9}, 3.1\cdot10^{-10}, 2.5\cdot10^{-10}, 2.0\cdot10^{-10}, 9.\cdot10^{-11}, 
  7.\cdot10^{-11}, 6.\cdot10^{-11}, 2.\cdot10^{-11}, \nn\\&2.\cdot10^{-11}, 1.\cdot10^{-11}, \}, \nn\\&\{7.919\cdot10^{-8}, 
  2.69\cdot10^{-9}, 1.3\cdot10^{-10}, 3.\cdot10^{-11}, 1.\cdot10^{-11}\}\nn
  \end{align}
   %%%%%%%%%%%%%%%%%%%%%%%%%%%%%%%%%%%%%%%%%%%%%%%%%%%%%%%%%%%%%%%%%%%%%%%%%%%%%%%%%%%
  

\bigskip


\begin{thebibliography}{[FJKLM]}


  \bibitem{bk} H.E.~Boos and V. E.~ Korepin.
  \newblock
Quantum spin chains and Riemann zeta function with odd arguments,
{\em J. Phys. A} {\bf 34} (2001) 5311-5316


\bibitem{XXX} 
H.~Boos, M.~Jimbo, T.~Miwa, F.~Smirnov and Y.~Takeyama. 
\newblock A recursion formula for the correlation 
functions of an inhomogeneous XXX model,
{\em Algebra i Analiz}	{\bf 17} (2005) 115-159

\bibitem{takahashi} J.~Sato, M.~ Shiroishi, M.~Takahashi.
\newblock
Exact evaluation of density matrix elements for the Heisenberg chain
{\it J.Stat.Mech.} {\bf 0612} (2006) P12017


\bibitem{HGSIII}
M.~Jimbo, T.~Miwa, and F.~Smirnov.
\newblock Hidden {Grassmann} structure in the {XXZ} model {III}: {Introducing
  Matsubara} direction.
\newblock {\em J. Phys. A} {\bf 42} (2009)  304018 (31pp)

\bibitem{corr} Ph. Di Francesco, F.~Smirnov. OPE for XXX, 
arXiv:1711.04123

\bibitem{FST}  L.D.~Faddeev, E.K.~Sklyanin, L.A.~Takhtajan.
\newblock
The quantum inverse problem method {\it  Theoretical and Mathematical Physics} {\bf 40} (1980) 688-711

\bibitem{BIK} V.E.~Korepin, N.M.~Bogoliubov, A.G.~Izergin \newblock Quantum inverse scattering method and
correlation functions {\it Cambridge University Press} (1993) 

\bibitem{Slavnov} N.A.~Slavnov.
\newblock
Calculation of scalar products of wave functions and form factors in the framework of the alcebraic Bethe ansatz
{\it  Theoretical and Mathematical Physics} {\bf 79} (1989) 502-508


\bibitem{J-lukyanov}V.~E.~ Korepin, S.~Lukyanov, Y. ~Nishiyama, M. ~Shiroishi. \newblock{Asymptotic Behavior of the Emptiness Formation Probability in the Critical Phase of XXZ Spin Chain}
{\it  Phys.Lett.} {\bf A312} (2003) 21-26



\bibitem{cft} 
C.~Holzhey C, F.~Larsen,  F.~Wilczek.  Geometric and renormalized entropy in conformal field theory {\it Nucl. Phys. B} {\bf 424} (1994 )  443-467

\bibitem{klumper}
A.~Kl\"umper.Thermodynamics of the anisotropic spin-1/2 Heisenberg chain and related quantum chains {\it Zeitschrift f\"ur Physik B Condensed Matter}
{\bf 91}(1993) 507Ð519

\bibitem{bgks}
H.~Boos, F.~G\"ohmann, A.~Kl\"umper and J.~Suzuki. \newblock
Factorization of the nite temperature
correlation functions of the XXZ chain in a magnetic eld" {\em J. Phys. A} 40 (2007) 10699-
10727


\bibitem{HGSIV}
 H.Boos, M.~Jimbo, T.~Miwa, F. Smirnov.
\newblock Hidden {Grassmann} structure in the {XXZ} model {IV}: {CFT} limit.
\newblock {\em Commun. Math. Phys.}  {\bf 299} (2010) 825--866


\bibitem{MM} {M.~ Mori, M.~Sugihara.
The double-exponential transformation in numerical analysis
{\em Journal of Computational and Applied Mathematics} {\bf 127}  (2001) 287--296}

\bibitem{ooura}{T.~Ooura.
Double exponential quadratures for various kinds of integral,
Talk given at
Second International ACCA-JP/UK Workshop, January 19, (2016)
Kyoto University}

\bibitem{korepin}
V.E.~Korepin \newblock
{\em Physical Review Letters} {\bf}92, issue 9, electronic identifier 096402, 05 March 2004



\bibitem{cardy} P.~ Calabrese, J.~Cardy.
Entanglement entropy and conformal field theory
{\em J.Phys.A} {\bf 42} (2009)  504005 -504036


\end{thebibliography}
 \end{document}